# Origin of the low critical observing temperature of the quantum anomalous Hall effect in V-doped (Bi, Sb)$_2$Te$_3$ film

W. Li[1,**], M. Claassen[1,**], Cui-Zu Chang[2,**], B. Moritz[1], T. Jia[1,3], C. Zhang[1], S. Rebec[1,3], J. J. Lee[1,3], M. Hashimoto[4], D.-H. Lu[4], R. G. Moore[1], J. S. Moodera[2,5], T.P. Devereaux[1,3,*], & Z.-X. Shen[1,3,*]

[1]Stanford Institute for Materials and Energy Sciences, SLAC National Accelerator Laboratory and Stanford University, Menlo Park, California 94025, USA

[2]Francis Bitter Magnet Lab, Massachusetts Institute of Technology, Cambridge, MA 02139, USA

[3]Departments of Physics and Applied Physics, and Geballe Laboratory for Advanced Materials, Stanford University, Stanford, California 94305, USA

[4]Stanford Synchrotron Radiation Lightsource, SLAC National Accelerator Laboratory, Menlo Park, California 94025, USA

[5]Department of Physics, Massachusetts Institute of Technology, Cambridge, MA 02139, USA

*To whom correspondence should be addressed: zxshen@stanford.edu, tpd@stanford.edu

** These authors contribute equally to this work

The experimental realization of the quantum anomalous Hall (QAH) effect in magnetically-doped (Bi, Sb)$_2$Te$_3$ films stands out as a landmark of modern condensed matter physics. However, ultra-low temperatures down to few tens of mK are needed to reach the quantization of Hall resistance, which is two orders of magnitude lower than the ferromagnetic phase transition temperature of the films. Here, we systematically study the band structure of V-doped (Bi, Sb)$_2$Te$_3$ thin films by angle-resolved photoemission spectroscopy (ARPES) and show unambiguously that the bulk valence band (BVB) maximum lies higher in energy than the surface state Dirac point. Our results demonstrate clear evidence that localization of BVB carriers plays an active role and can account for the temperature discrepancy.

The QAH effect[1-12], characterized by its dissipationless spin-polarized chiral edge states at zero magnetic field, provides opportunities for future applications in low-energy-dissipation electronic and spintronics devices. Conceptually, the QAH effect in magnetically-doped topological insulator (TI) films is expected to arise from the interplay of magnetic-dopant exchange coupling and hybridization of the top and bottom surface Dirac cones, with both contributions inducing a mass gap at the Dirac point (DP)[4, 6]. If the exchange-coupling mediated Zeeman term dominates, the hybridized surface bands invert only for one spin orientation and induce a single chiral edge mode that spans the '2D bulk" gap of the thin film at the DP (Fig. 1c). This mechanism of exchange-inverted surface states (QAH-ESS) stands in contrast to the conventional quantum Hall (QH) effect which relies on Landau quantization of the electronic motion by an external magnetic field. While ferromagnetism sets in at Curie temperatures on the order of tens of Kelvins, the relevant transport energy scales are ultimately limited by the effective size of the band gap that is set by the exchange splitting at the Dirac point (DP). Therefore, in materials where the three dimensional (3D) bulk TI valence band maximum (VBM) lies well below the DP (Fig. 1b), the QAH-ESS effect upon magnetic doping should be observable below temperatures whose scale is set by the lesser of the Curie temperature and the exchange-induced Dirac gap. Indeed, first-principles calculations for magnetically-doped $Bi_2Se_3$ indicate that ferromagnetic ordering would induce an exchange splitting of approximately 40meV for 5 quintuple layers (QLs)[6], suggesting a significantly higher QAH temperature than observed in experiment for both Cr- and V-doped $(Bi, Sb)_2Te_3$ films.

However, the relative energy of the VBM with respect to the DP ($\Delta$) remains unclear in magnetically-doped alloyed $(Bi, Sb)_2Te_3$ films: as shown in Fig. 1, previous studies reveal positive $\Delta$ in $Bi_2Te_3$[13, 14] and negative $\Delta$ in $Sb_2Te_3$[13, 15], with first-principles studies for V doping indicating a separate *d*-electron impurity band located at the Fermi level[6]. A detailed determination is crucial to understand the nature of the quantized Hall signatures: If Δ is a small negative value, whose magnitude is comparable to the exchange-induced gap, the total effective gap may be reduced. Conversely, if Δ>0 the effective gap should disappear entirely, naively suggesting an absence of quantized Hall signatures in these samples due to the transport contributions from the BVB (Fig. 1c).

**Results**

We use ARPES to reveal the band structure of a $(Bi_{0.29}Sb_{0.71})_{1.89}V_{0.11}Te_3$ QAH film[11] (Fig. 2a and Fig. S1). The QAH thin films were grown by custom-built molecular beam epitaxy (MBE), and stem from the same batch of films previously used to observe the QAH effect in transport[11]. Details of the sample can be found in the Methods section and Ref. 11. The DP is 54 meV below the Fermi energy ($E_F$) (Supplementary Note 1 and Fig. 2a). No ferromagnetic exchange gap is resolved near the DP down to 7 K (while the Curie temperature $T_c$ of the compound is 19 K), which is expected due to the limited energy resolution. Consistent with the theoretical calculation[13], the BVB (marked by the yellow arrows) along the Γ–M direction (Fig. S1a) is closer in energy to the DP than that along the Γ–K direction (Fig. S1b). However, it is still challenging to decide based on this spectrum whether the BVB crosses the energy of the DP ($E_D$): The intensity of a band in ARPES spectra could be weak due to the matrix element effect, and we therefore need a systematic way to accurately determine the location of the BVB maximum.

The capability of mapping the constant energy contours (CEC) by ARPES provides an ideal way to explore the band structure of a material along different directions in momentum space, while at the same time helping to overcome matrix element effects. We demonstrate this with a Bi$_2$Te$_3$ film. With a photon energy of 20.5 eV, the VBM of a five QL Bi$_2$Te$_3$ film is highlighted, showing the DP located deeply below the VBM [the inset of Fig. 2b], consistent with literature[14]. The first column of Fig. 2b shows data for a photon energy of 29.5 eV that highlights the surface state band (SSB) for more accurate determination of DP energy. Due to a weaker intensity from the BVB, it remains unclear in the energy dispersion along Γ-M-K whether the VBM is higher than DP. However, in the CEC map at $E_D$, the BVB still manifests itself as a six-fold flower-shaped structure surrounding the Dirac cone (see the CEC at $E_D$ = -376 meV in the second column and the schematic in the third column of Fig. 2b). The petals of the flower are along the Γ–M direction. Such a flower-shaped CEC hence acts as a signature of the BVB in this family of TIs[14-16]. Inspired by results on Bi$_2$Te$_3$, we carefully measured the CEC maps of the V-doped (Bi, Sb)$_2$Te$_3$ film. As shown in Fig. 2a, the SSB induces a small circular Fermi surface at $E_F$. The flower-shaped CEC around the DP is already developed at $E_D \sim$ -54 meV. At higher binding energy, the flower-shaped structure gradually expands. The coexistence of the DP and the flower-shaped structure in CEC provides direct evidence for the overlap of the DP and the BVB in this system.

Figure 3b shows the CEC map of the V-doped (Bi, Sb)$_2$Te$_3$ film at $E_D$. As discussed above, the flower-shaped structure around the DP arises from the BVB. Along cut 1 (see Fig. S4 as well as the three-dimensional illustration of the Dirac cone in Fig. 3a), the VBM is hard to determine due to the matrix element effect. A better angle to reveal the VBM is slightly off the Γ point, which directly cuts the two petals of the CEC (cut 2). As schematically shown in Fig. 3b, one expects a spectrum composed of petals of the flower-shaped CECs (the yellow part) along the energy direction, centered with the intensity from SSB (the blue part). Fig. 3c shows the band dispersion along cut 2, where two branches of the BVB clearly cross the $E_D$ [also see the momentum distribution curve (MDC) at $E_D$ that reveals a double peak structure]. Therefore, we unambiguously demonstrate that the VBM lies above $E_D$.

**Discussion**

Our observations clearly establish the absence of a thin film bulk band gap at the DP energy that would be expected for the QAH-ESS. The consequences are two-fold: on one hand, RKKY interaction through BVB states can complement the V core-level van-Vleck contribution[17] to ferromagnetism, to explain the observed robust magnetic order of V moments. Second and more importantly, the absence of a bulk band gap should entail a bulk contribution to transport, precluding the QAH effect. An essential next step is therefore to reconcile this result with the reported QAH phenomena and ultra-low observing temperatures of few tens of mK, orders of magnitude lower than the Curie temperature or the expected exchange-induced gap[6, 11, 18, 19]. Importantly, in the QAH-ESS scenario, a clear transport signature of the chiral edge mode and associated quantization of $\sigma_{xy}$ necessitates that the BVB carriers are absent in transport.

As 2D electrons necessarily localize in the absence of a magnetic field and spin-orbit coupling, the absence of BVB electrons in transport points, at first glance, to impurity-driven 2D Anderson localization (Fig. 3d and Fig. S5). Such a picture holds only if the thin film bulk falls into the unitary universality class. However, strong spin-orbit coupling expected for a TI should place the bulk in the symplectic class, leading to anti-localization. The question then remains whether time reversal symmetry breaking due to magnetic impurities can still lead to localization. In principle, the QAH effect then should be determined by the temperature dependence of the localization length. If the insulating 2D bulk in the QAH phase indeed follows from Anderson localization, then samples with higher magnetic dopant concentration should naively suppress

BVB contributions to $\sigma_{xx}$ and push quantized Hall signatures to higher temperatures, up to the point where the disorder bandwidth is comparable to the exchange-induced gap.

Another potential contribution follows from considering the internal magnetic field that arises from the ferromagnetic moments, ~13 mT for a fully-polarized domain of ~ 1.5 $\mu_B$/V at 5.5% doping. In the absence of spin-orbit coupling, a BVB effective mass ~ 0.18 $m_0$ suggests a magnetization-induced cyclotron frequency $\omega_c$ ~ 100 mK, above the temperature of observed quantized Hall signatures but well below the Curie temperature, and increasing to $B_{int}$ ~ 200mT and $\omega_c$ ~2K for the penta-layer QAH samples[20]. In theory, the internal magnetic field affects BVB electrons in two ways: First, the induced cyclotron gap can help quench bulk transport $\sigma_{xx}$, introducing the magnetic length as an extra scale in the localization problem. Second, extended states of the BVB Landau levels could conceivably provide a complementary QAH mechanism of a magnetic field driven "quantum Hall" $\sigma_{xy}$ contribution. The latter scenario would require that no simultaneous exchange-induced topological band inversion takes place for the SSBs and is rather unlikely due to the exceedingly low mobility of the samples. In experiments on V and Cr doped films, $B_{int}$ is much smaller than the coercivity field: a clear counter-indication then comes from magnetic hysteresis curves in which $\sigma_{xy}$ does not disappear when the external field sweeps across -$B_{int}$, cancelling the total magnetic field[11].

A third possibility for the absence of a bulk band contribution to transport arises if the chemical potential $\mu$ of the gated sample, tuned to the QAH regime, lies above the BVB maximum, inside the upper SSB. In this case, the BVB remains inert. Therefore, disorder and internal magnetic field must instead localize SSB electrons. While the TI surface state exhibits anti-localization in the absence of time-reversal symmetry breaking, a transition to localization occurs as a function of $\mu$-$E_D$ and magnetic scattering[21]. This scenario benefits from modulation doping of the penta-layer samples[20]: as the Cr dopants are concentrated near the film surfaces, scattering is enhanced for the SSBs with peak amplitudes at the top and bottom surface, further aiding localization. On the other hand, the QAH edge modes would need to extend beyond the range of the exchange-induced gap, far above $E_D$. However, given the Curie temperature and density-functional-theory predictions of the QAH edge mode dispersion[22], this is unlikely.

In summary, among the three scenarios discussed, localization of the BVB appears to be the most likely explanation for the much lower-than-expected temperatures of the QAH effect. Therefore, the solution might be searching for a system in which the Dirac point is well separated from BVB to avoid the bulk carriers, and not too far from $E_F$ for easily gate tuning. Very recently, a bulk insulating TI has been realized in Sn-doped $Bi_{1.1}Sb_{0.9}Te_2S$ single crystals[23], which might be a good platform for further study of QAH. However, open questions remain. For example, recently-reported experimental results on thicker (10QL) films with negligible top and bottom SSB hybridization, which should naively reduce the temperature scale, nevertheless display the QAH effect at similar temperature[9, 10, 12]. Furthermore, questions regarding the importance of localization come from Cr doped penta-layer QAH samples[20], where modulation doping conceivably reduces disorder in the thin film bulk, counteracting localization of a BVB while at the same time pushing QAH signatures to elevated temperatures. Given its potential importance for the QAH effect, a better understanding of the Anderson localization problem due to interplay of disorder, spin-orbit coupling and the internal magnetic field will be essential. Given that the internal magnetization-induced cyclotron frequency $\omega_c \sim 100$ mK is well above the temperature of observed quantized Hall signatures, it is conceivable that impurity-driven Anderson localization for the bulk valence band and internal magnetic field induced Landau quantization may cooperate with each other to suppress bulk transport and while retaining chiral edge conduction. As transport properties are analogous for both V and Cr doping, it needs to be checked whether similar conclusions about the position of the BVB relative to $E_D$ pertain to Cr doped thin films. Finally, direct detection of the exchange-mediated gap of the SSBs as well as a systematic study of the role of disorder using magnetic and non-magnetic dopants could further shed light on the microscopic mechanism that ultimately gives rise to the observed quantized transport signatures, as well as to delineate future directions towards a higher-temperature QAH effect.

To recap, our presented results clearly establish the absence of a thin film bulk gap at $E_D$, which suggests that the observed of QAH is a consequence of a chiral edge state in the presence of bulk carriers. This points towards possible origins why the temperature of quantized transport signatures is significantly lower than expected for an exchange mediated SSB inversion of topological insulator.

**Methods**

The 4QL V-doped (Bi, Sb)$_2$Te$_3$ QAH thin films were grown by custom-built molecular beam epitaxy (MBE) with a base pressure better than $5\times10^{-10}$ Torr[11]. 5 nm Te was evaporated on the top as a capping layer for *ex-situ* ARPES measurements. The Te capping layers were removed by heating the films with a filament behind the sample holder in ultra-high vacuum (UHV) chamber before ARPES measurements. The de-capping procedure was monitored by RHEED and the de-capping temperature was optimized (Supplementary Note 2). The ARPES measurements were performed at the Stanford Synchrotron Radiation Lightsource (SSRL) Beamline 5-4 at 22 K. The photon energies of 29.5 eV and 20.5 eV were selected to highlight the SSB and the BVB, respectively. The energy resolution was set at 10.5 meV.

# References

1. Nagaosa, N, Sinova, J & Onoda, S *et al*. Anomalous Hall effect. *Rev. Mod. Phys.* **82**, 1539–92 (2010).

2. Onoda, M & Nagaosa, N. Quantized anomalous Hall effect in two-dimensional ferromagnets: quantum Hall effect in metals. *Phys. Rev. Lett.* **90**, 206601 (2003).

3. Qi, X. L., Wu, Y. S. & Zhang, S. C. Topological quantization of the spin Hall effect in two dimensional paramagnetic semiconductors. *Phys. Rev. B* **74**, 085308 (2006)

4. Qi, X. L., Hughes, T. L. & Zhang, S. C. Topological field theory of time-reversal invariant insulators. *Phys. Rev. B* **78**, 195424 (2008).

5. Liu, C. X., Qi, X. L., Dai, X., Fang, Z. & Zhang, S. C. Quantum anomalous hall effect in $Hg_{1-y}Mn_yTe$ quantum wells. *Phys. Rev. Lett.* **101**, 146802 (2008).

6. Yu, R. *et al*. Quantized anomalous Hall effect in magnetic topological insulators. *Science* **329**, 61 (2010).

7. Nomura, K. & Nagaosa, N. Surface-quantized anomalous Hall current and the magnetoelectric effect in magnetically disordered topological insulators. *Phys. Rev. Lett.* **106**, 166802 (2011).

8. Chang, C. Z. *et al*. Experimental observation of the quantum anomalous Hall effect in a magnetic topological insulator. *Science* **340**, 167–170 (2013).

9. Kou, X. *et al*. Scale-invariant quantum anomalous Hall effect in magnetic topological insulators beyond two-dimensional limit. *Phys. Rev. Lett.* **113**, 137201 (2014).

10. Checkelsky, J. G. *et al*. Trajectory of the anomalous Hall effect toward the quantized state in a ferromagnetic topological insulator. *Nature Phys.* **10**, 731–736 (2014).

11. Chang, C. Z. *et al*. High-precision realization of robust quantum anomalous Hall state in a hard ferromagnetic topological insulator. *Nature Mater.* **14**, 473–477 (2015).

12. Bestwick, A. J. *et al*. Precise quantization of the anomalous Hall effect near zero magnetic field. *Phys. Rev. Lett.* **114**, 187201 (2015).

**Acknowledgements**

We would like to acknowledge very helpful discussions with Xi Dai, Shoucheng Zhang, Xiaoliang Qi, Naoto Nagaosa and Dunghai Lee. ARPES experiments were supported by the Office of Basic Energy Science, Division of Materials Science, and were performed at the Stanford Synchrotron Radiation Lightsource, which is operated by the Office of Basic Energy Sciences, U.S. Department of Energy.

**Author contributions**

W. L. and Z.-X. S. conceived the project. W. L. took the ARPES measurements with assistance from T. J., C. Z., S. R. and J. J. L.. C.-Z. C. grew the samples. W. L., M. C. and B. M. analyzed the data. M. H., R. G. M. and D.-H. L. assisted in ARPES measurements at Stanford Synchrotron Radiation Lightsource. J. S. M., T. P. D., and Z.-X. S. supervised the project. W. L., M. C., C.-Z. C., B.M., T.P. D. and Z.-X. S. wrote the paper with input from all coauthors.

**Additional Information**

Competing financial interests

The author(s) declare no competing financial interests.

**Figure Captions**

**Figure 1.** Schematic showing the relative position of the DP with respect to the VBM in $Bi_2Te_3$, $Sb_2Te_3$ and QAH samples, respectively. The dashed lines indicate the position of the DP and the VBM of the corresponding bulk materials. The red and blue colors of the surface bands denote the even and odd parities, respectively. Due to the finite thickness of all measured films, a tiny hybridized gap opens at the DP as shown in the schematic. In **c**, exchange coupling with ferromagnetically-ordered dopants induces a Zeeman splitting of the surface bands, whose spin orientations are denoted schematically via dashed and solid lines. A pair of inverted surface bands then appears when the exchange splitting surpasses the hybridization gap, inducing a single chiral edge mode (the green solid line) that spans the thin film gap of the surface bands.

**Figure 2.** Signature of the BVB in CEC maps. Band dispersion of a 4QL $(Bi_{0.29}Sb_{0.71})_{1.89}V_{0.11}Te_3$ QAH sample along a cut from M to Γ is shown in the first column of **a**, however the relative position between the DP and the VBM needs further determination. As shown in **b**, the DP of a reference sample $Bi_2Te_3$ locates deeply below the VBM. The inset highlights the VB with photon energy of 20.5 eV. The flower-shaped structure surrounding the Dirac cone in the CEC maps in the second column indicates the existence of valence band (see also the yellow part of BVB in the schematic). In analogy to $Bi_2Te_3$, in **a** the BVB (flower-shaped structure) could be observed near the DP in the CEC map, which is strong evidence of the overlap of the VBM and the DP in QAH system.

**Figure 3.** Direct observation of the overlap of the VBM and the DP. **a**, Three-dimensional illustration of the Dirac cone in QAH system. **b**, CEC map at the energy of the DP. The solid yellow lines across the Brillouin zone indicate the cut directions of the band dispersion spectra in **c** and **Fig. S4**. **c**, High-resolution band dispersion along cut 2. Two symmetric branches of BVB

centered at $k = 0$ cross the $E_D$. Photon energy of 20.5 eV is chosen to highlight the VB. As shown in **b**, the BVB in cut 2 connects the petals of the flower-shaped CECs with different energies. **d**, Complementary approaches to the bulk localization problem: Anderson localization of the BVB with strong spin orbit coupling can lead to an insulating 2D bulk at low temperatures only if the sample lies in the unitary symmetry class. Conversely, the internal magnetic field of V moment can quantize electron motion in the BVB to insulate the thin film bulk. The QAH sample hosts counter-propagating chiral edge states at the thin film edges.

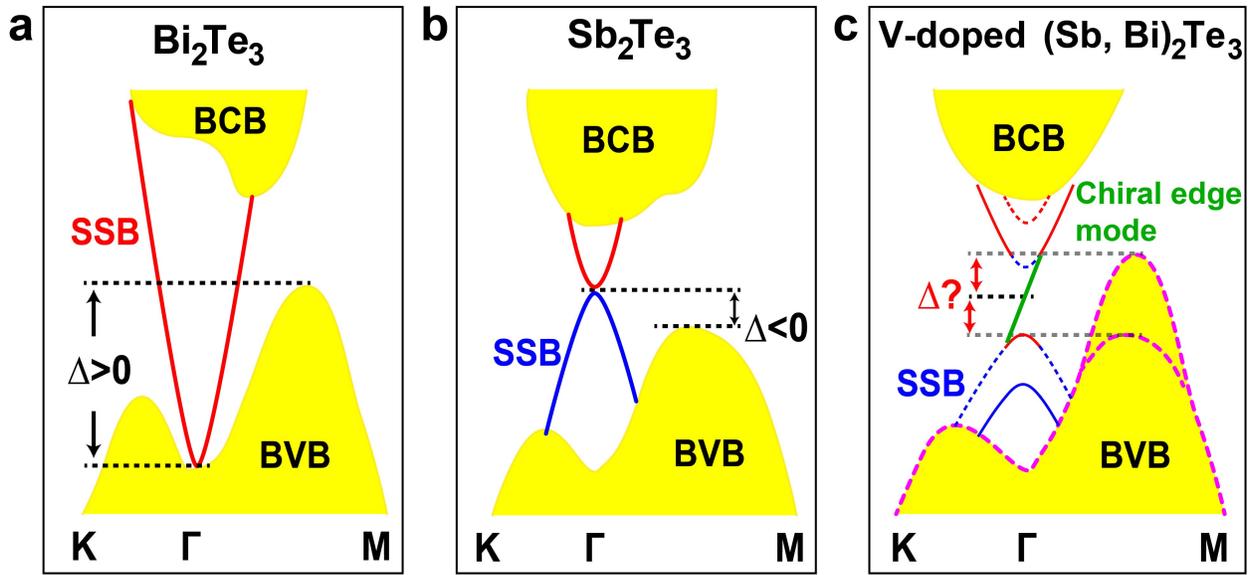

Figure 1

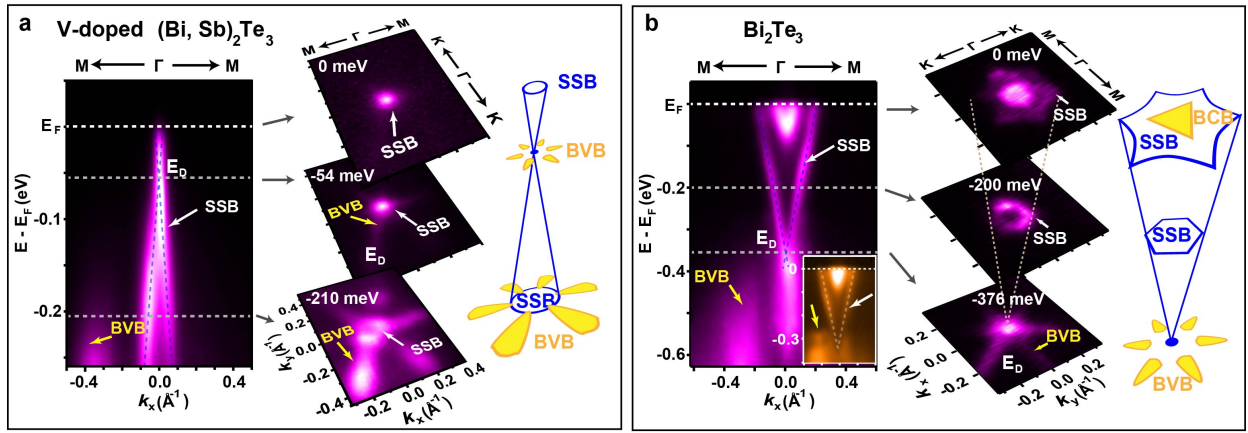

**Figure 2**

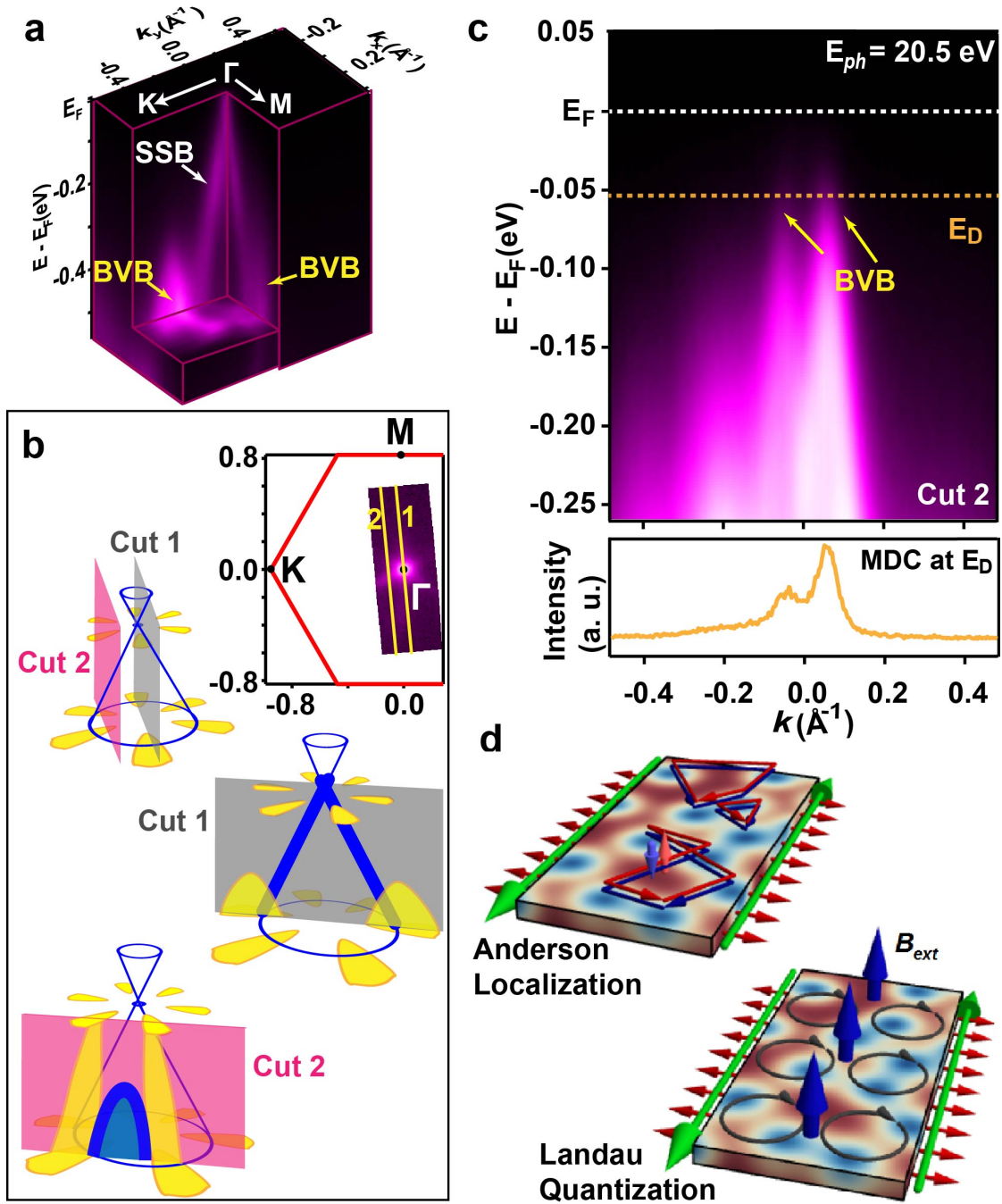

**Figure 3**